\documentclass[nofootinbib,preprint,double-spaced]{revtex4}%
\usepackage{amssymb}
\usepackage{amsfonts}
\usepackage{amsmath}
\usepackage{graphicx}%
\begin{document}
\title{\normalsize Exact solutions for the Einstein-Gauss-Bonnet theory in five dimensions:\\Black
holes, wormholes and spacetime horns}
\author{Gustavo Dotti$^{1}$, Julio Oliva$^{2,3}$, and Ricardo Troncoso$^{3}$}
\affiliation{$^{1}$Facultad de Matem\'{a}tica, Astronom\'{\i}a y F\'{\i}sica, Universidad
Nacional de C\'{o}rdoba, Ciudad Universitaria, (5000) C\'{o}rdoba, Argentina. }
\affiliation{$^{2}$Departamento de F\'{\i}sica, Universidad de Concepci\'{o}n, Casilla,
160-C, Concepci\'{o}n, Chile.}
\affiliation{$^{3}$Centro de Estudios Cient\'{\i}ficos (CECS), Casilla 1469, Valdivia, Chile.}

\begin{abstract}
An exhaustive classification of certain class of static solutions for the
five-dimensional Einstein-Gauss-Bonnet theory in vacuum is presented. The
class of metrics under consideration is such that the spacelike section is a
warped product of the real line with a nontrivial base manifold. It is shown
that for generic values of the coupling constants the base manifold must be
necessarily of constant curvature, and the solution reduces to the topological
extension of the Boulware-Deser metric. It is also shown that the base
manifold admits a wider class of geometries for the special case when the
Gauss-Bonnet coupling is properly tuned in terms of the cosmological and
Newton constants. This freedom in the metric at the boundary, which determines
the base manifold, allows the existence of three main branches of geometries
in the bulk. For negative cosmological constant, if the boundary metric is
such that the base manifold is arbitrary, but fixed, the solution describes
black holes whose horizon geometry inherits the metric of the base manifold.
If the base manifold possesses a negative constant Ricci scalar, two different
kinds of wormholes in vacuum are obtained. For base manifolds with vanishing
Ricci scalar, a different class of solutions appears resembling \textquotedblleft spacetime
horns". There is also a special case for which, if the base manifold is of
constant curvature, due to certain class of degeneration of the field
equations, the metric admits an arbitrary redshift function. For wormholes and
spacetime horns, there are regions for which the gravitational and centrifugal
forces point towards the same direction. All these solutions have finite
Euclidean action, which reduces to the free energy in the case of black holes,
and vanishes in the other cases. The mass is also obtained from a surface integral.

\textmd{\textit{\textbf{Electronic addresses:} gdotti-at-famaf.unc.edu.ar,
juliooliva-at-cecs.cl, ratron-at-cecs.cl}}

\end{abstract}
\maketitle
\tableofcontents



\section{Introduction}

\label{intro}

According to the basic principles of General Relativity, higher dimensional
gravity is described by theories containing higher powers of the curvature
\cite{Lovelock}. In five dimensions, the most general theory leading to second
order field equations for the metric is the so-called Einstein-Gauss-Bonnet
theory, which contains quadratic powers of the curvature. The pure gravity
action is given by
\begin{equation}
I=\kappa\int d^{5}x\sqrt{g}\left(  R-2\Lambda+\alpha\left(  R^{2}-4R_{\mu\nu
}R^{\mu\nu}+R_{\alpha\beta\gamma\delta}R^{\alpha\beta\gamma\delta}\right)
\right)  \ ,\label{Itensor}%
\end{equation}
where $\kappa$ is related to the Newton constant, $\Lambda$ to the
cosmological term, and $\alpha$ is the Gauss-Bonnet coupling. For later
convenience, it is useful to express the action (\ref{Itensor}) in terms of
differential forms as%
\begin{equation}
I=\int\!\epsilon_{abcde}\left(  \!\alpha_{2}R^{ab}R^{cd}\!+\!\alpha_{1}%
R^{ab}e^{c}e^{d}\!+\!\alpha_{0}e^{a}e^{b}e^{c}e^{d}\!\right)  \!e^{e}%
\ ,\label{action}%
\end{equation}
where $R^{ab}=d\omega^{ab}+\omega_{\text{ \ }f}^{a}\omega^{fb}$ is the
curvature $2$-form for the spin connection $\omega^{ab}=\omega_{\ \mu}%
^{ab}dx^{\mu}$, $e^{a}=e_{\ \mu}^{a}dx^{\mu}$ is the vielbein and the wedge
product is understood \footnote{The relationship between the constants
appearing in Eqs (\ref{Itensor}) and (\ref{action}) is given by $\alpha
=\frac{\alpha_{2}}{6\alpha_{1}}$, $\Lambda=10\frac{\alpha_{0}}{\alpha_{1}}$,
$\kappa=-6\alpha_{1}$ .\textbf{\ }}. For a metric connection with vanishing
torsion, the field equations from (\ref{action}) read%
\begin{equation}
\mathcal{E}_{a}:=\epsilon_{abcde}\!\left(  \alpha_{2}R^{bc}R^{de}%
\!+3\alpha_{1}R^{bc}e^{d}e^{e}\!+5\alpha_{0}e^{b}e^{c}e^{d}e^{e}\right)
\!=0\ .\label{eom}%
\end{equation}

The kind of spacetimes we are interested in have static metrics of the form
\begin{equation}
ds^{2}=-f^{2}\left(  r\right)  dt^{2}+\frac{dr^{2}}{g^{2}\left(  r\right)
}+r^{2}d\Sigma_{3}^{2}\ ,\label{Ansatz}%
\end{equation}
where $d\Sigma_{3}^{2}$ is the line element of a three-dimensional manifold
$\Sigma_{3}$ that we call the \textquotedblleft base manifold". Note that
$\partial/\partial t$ is a timelike Killing vector field, orthogonal to
4-manifolds that are a warped product of ${\mathbb{R}}$ with the base manifold
$\Sigma_{3}$.

If the Gauss-Bonnet coupling $\alpha_{2}$ vanishes, General Relativity with a
cosmological constant is recovered. In this case the equations force the base
manifold to be of constant curvature $\gamma$ (which can be normalized to
$\gamma=\pm1$ or zero) and \footnote{The four dimensional case was discussed
previously in \cite{ehtop}, \cite{ehtop1}, \cite{ehtop2}.}\ \cite{Birmingham}%
\begin{equation}
f^{2}=g^{2}=\gamma-\frac{\mu}{r^{2}}-\frac{5}{3}\frac{\alpha_{0}}{\alpha_{1}%
}r^{2}~,
\end{equation}
If $\gamma=1$, i.e., for $\Sigma_{3}=S^{3}$, the Schwarzschild-anti-de Sitter
solution is recovered.\newline For spacetime dimensions higher than five, the
equations of General Relativity do not impose the condition that the base
manifold be of constant curvature. In fact, \emph{any} Einstein base manifold
is allowed \cite{gh}. For nonzero $\alpha_{2}$, however, the presence of the
Gauss-Bonnet term restricts the geometry of an Einstein base manifold by
imposing conditions on its Weyl tensor \cite{Dotti-Gleiser}.

In this work we restrict ourselves to five dimensions \emph{without assuming
any a priori condition on the base manifold} in the ansatz (\ref{Ansatz}). We
show that in five dimensions, the presence of the Gauss-Bonnet term permits to
relax the allowed geometries for the base manifold $\Sigma_{3}$, so that the
whole structure of the five-dimensional metric turns out to be sensitive to
the geometry of the base manifold. More precisely, it is shown that solutions
of the form (\ref{Ansatz}) can be classified in the following way:\newline

\noindent$\mathbf{\circ}$\textbf{\ (i) Generic class:} For generic
coefficients, i.e., for arbitrary $\alpha_{0}$, $\alpha_{1}$, $\alpha_{2}$,
the line element (\ref{Ansatz}) solves the Einstein-Gauss-Bonnet field
equations provided the base manifold $\Sigma_{3}$ is of constant curvature
$\gamma$ (that we normalize to $\pm1,0$) and
\begin{equation}
f^{2}=g^{2}\left(  r\right)  =\gamma+\frac{3}{2}\frac{\alpha_{1}}{\alpha_{2}%
}r^{2}\left[  1\pm\sqrt{\left(  1-\frac{20}{9}\frac{\alpha_{2}\alpha_{0}%
}{\alpha_{1}^{2}}\right)  +\frac{\mu}{r^{4}}}\;\right]
\ ,\label{gcuadradogenerico}%
\end{equation}
where $\mu$ is an integration constant \cite{Cai}. In the spherically
symmetric case, (\ref{gcuadradogenerico}) reduces to the well known
Boulware-Deser solution \cite{BD}.\newline

\noindent$\mathbf{\circ}$\textbf{\ (ii) Special class:} In the special case
where the Gauss-Bonnet coupling is given by%
\begin{equation}
\alpha_{2}=\frac{9}{20}\frac{\alpha_{1}^{2}}{\alpha_{0}}\ ,\label{tuneo}%
\end{equation}
the theory possesses a unique maximally symmetric vacuum \cite{BH-Scan}, and
the Lagrangian can be written as a Chern-Simons form \cite{Chamseddine}. The
solution set splits into three main branches according to the geometry of the
base manifold $\Sigma_{3}$:\newline

\noindent$\cdot$\textit{\ (ii.a) Black holes:}\newline

These are solutions of the form (\ref{Ansatz}) with
\begin{equation}
f^{2}=g^{2}=\sigma r^{2}-\mu~,\;\;\sigma:=\frac{10}{3}\frac{\alpha_{0}}%
{\alpha_{1}}\label{gbhspecial1}%
\end{equation}
($\mu$ an integration constant). Their peculiarity is that with the above
choice of $f$ and $g$, \emph{any} (fixed) base manifold $\Sigma_{3}$ solves
the field equations. Note that for negative cosmological constant $\left(
\sigma>0\right)  $ this solution describes a black hole \cite{Cai-Soh},
\cite{ATZ}, which in the case of spherical symmetry, reduces to the one found
in \cite{BD}, \cite{BTZ}.\newline

\noindent$\cdot$\textit{\ (ii.b1) Wormholes and spacetime horns:}\newline

For base manifolds $\Sigma_{3}$ of constant nonvanishing Ricci scalar,
$\tilde{R}=6\gamma$, the metric (\ref{Ansatz}) with
\begin{align}
f^{2}(r)  &  =\left(  \sqrt{\sigma}r+a\sqrt{\sigma r^{2}+\gamma}\right)
^{2}~,\label{fcuaworm}\\
g^{2}\left(  r\right)   &  =\sigma r^{2}+\gamma~,\label{gcuaesp}%
\end{align}
($a$ is an integration constant) is a solution of the field equations. In this
case, there are three subbranches determined by $|a|>1$, $|a|<1$ or $|a|=1$.
It is simple to show that, for negative cosmological constant $\left(
\sigma>0\right)  $ and $\gamma=-1$, the solution with $|a|<1$ corresponds to
the wormhole in vacuum found in \cite{DOTwormhole}. The solution with $|a|=1$
and $\gamma=-1$ corresponds to a brand new wormhole in vacuum (See Section
III).\newline If the base manifold $\Sigma_{3}$ has \emph{vanishing} Ricci
scalar, i.e., $\tilde{R}=0$, it must be
\begin{align}
f^{2}(r)  &  =\left(  a\sqrt{\sigma}r+\frac{1}{\sqrt{\sigma}r}\right)
^{2}~,\label{f-horn}\\
g^{2}\left(  r\right)   &  =\sigma r^{2}~,\label{grr-horn}%
\end{align}
with $a$ an integration constant. If $\sigma>0$ and $a\geq0$ this solution
looks like a \textquotedblleft spacetime horn". If the base manifold is not
locally flat, there is a timelike naked singularity, but nevertheless the mass
of the solution vanishes and the Euclidean continuation has a finite action
(See Section IV).\newline

\noindent$\cdot$\textit{\ (ii.b2) Degeneracy:}\newline

If $\Sigma_{3}$ is of constant curvature, $\tilde{R}^{mn}=\gamma\tilde{e}%
^{m}\tilde{e}^{n}$, and $g^{2}$ given by Eq. (\ref{gcuaesp}), then the
function $f^{2}\left(  r\right)  $ is left undetermined by the field equations.

\bigskip

The organization of the paper is the following: in Section \ref{sols} we solve
the field equations and arrive at the classification outlined above, Section
\ref{geom} is devoted to describing the geometry of the solutions of the
special class, including some curious issues regarding the nontrivial behavior
of geodesics around wormholes and spacetime horns. The Euclidean continuation
of these solutions and the proof of the finiteness of their Euclidean action
is worked out in Section \ref{euc}. The mass of these solutions is computed
from surface integrals in Section \ref{mass}. Section \ref{conc} is devoted to
a discussion of our results, and some further comments\textbf{.}

\section{Exact solutions and their classification}

\label{sols}

In this Section we solve the field equations and arrive at the classification
outlined in Section \ref{intro}. This is done in two steps. We first solve the
constraint equation $\mathcal{E}_{0}=0$, and find two different cases: (i) a
solution which is valid for any Einstein-Gauss-Bonnet theory, (ii) a solution
that applies only to those theories satisfying (\ref{tuneo}). \newline In a
second step we solve the remaining field equations and complete the
classification of the solution set.\newline

The vielbein for the metric (\ref{Ansatz}) is chosen as%
\begin{equation}
e^{0}=fdt~,~e^{1}=g^{-1}dr~,~e^{m}=r\tilde{e}^{m}~,
\end{equation}
where $\tilde{e}^{m}$ stands for the vielbein on the base manifold, so that
the indices $m,n,p...$ run along $\Sigma_{3}$. The constraint equation
$\mathcal{E}_{0}=0$ then acquires the form
\begin{equation}
B_{0}\left(  r\right)  \tilde{R}+6A_{0}\left(  r\right)
=0~,\label{constraint}%
\end{equation}
where $\tilde{R}$ is the Ricci scalar of the base manifold, and%
\begin{align}
A_{0}  &  =20\alpha_{0}r^{4}\!-3\alpha_{1}r\left(  g^{2} r^{2} \right)
^{\prime}+\alpha_{2}\left(  g^{4}\right)  ^{\prime}r~,\\
B_{0}  &  =2r\left[  3\alpha_{1}r-\alpha_{2}\left(  g^{2}\right)  ^{\prime
}\right]  ~.
\end{align}
Since $\tilde{R}$ depends only on the base manifold coordinates,
Eq.(\ref{constraint}) implies that
\begin{equation}
A_{0}\left(  r\right)  =-\gamma B_{0}\left(  r\right)  ~,\label{apropbconst}%
\end{equation}
where $\gamma$ is a constant. Hence, the constraint reduces to%
\begin{equation}
\label{constrfact}%
\begin{cases}
B_{0}\left(  r\right)  \left(  \tilde{R}-6\gamma\right)  =0,\\
A_{0}\left(  r\right)  =-\gamma B_{0}\left(  r\right)
\end{cases}
\end{equation}
and implies that either

\bigskip

(i) the base manifold is of constant Ricci scalar $\tilde{R}=6\gamma$, or

(ii) $B_{0}=0$.

\bigskip

In case (i) the solution to (\ref{apropbconst}) is
\begin{equation}
g^{2}\left(  r\right)  =\gamma+\frac{3}{2}\frac{\alpha_{1}}{\alpha_{2}}%
r^{2}\left[  1\pm\sqrt{\left(  1-\frac{20}{9}\frac{\alpha_{2}\alpha_{0}%
}{\alpha_{1}^{2}}\right)  +\frac{\mu}{r^{4}}}\right]
~,\label{gcuagenconlambda}%
\end{equation}
($\mu$ is an integration constant). Since this solution holds for generic
values of $\alpha_{0},\alpha_{1}$ and $\alpha_{2}$ we call case (i) the
\emph{generic} branch.\newline

\noindent Case (ii), on the other hand, implies $A_{0}=B_{0}=0$ (see equation
(\ref{apropbconst})), and this system admits a solution only if the constants
of the theory are tuned as in (\ref{tuneo}), the solution being
\begin{equation}
g^{2}=\sigma r^{2}-\mu~, \;\; \sigma:=\frac{10}{3}\frac{ \alpha_{0}}%
{\alpha_{1}}\label{gbhspecial}%
\end{equation}
Note that in case (ii) the constraint equation does not impose any condition
on the base manifold.\newline

The radial equation $\mathcal{E}_{1}~=~0$, combined with the constraint in the
form $e^{0}\mathcal{E}_{0}-e^{1}\mathcal{E}_{1}=0$ reduces to
\begin{equation}
\left(  B_{0}\left(  r\right)  -B_{1}\left(  r\right)  \right)  \tilde
{R}+6\left(  A_{0}\left(  r\right)  -A_{1}\left(  r\right)  \right)
=0~,\label{eps1sinfac}%
\end{equation}
where
\begin{align*}
A_{1}\left(  r\right)   &  =2r\left[  10\alpha_{0}r^{3}-3\alpha_{1}%
g^{2}r-3\alpha_{1}g^{2}\frac{f^{\prime}}{f}r^{2}+2\alpha_{2}\frac{f^{\prime}%
}{f}g^{4}\right]  \ ,\\
B_{1}\left(  r\right)   &  =2r\left[  3\alpha_{1}r-2\alpha_{2}g^{2}%
\frac{f^{\prime}}{f}\right]  ~.
\end{align*}

Finally, the three ``angular" field equations $\mathcal{E}_{m}=0$ are
equivalent to the following three equations
\begin{equation}
B \left(  r\right)  \tilde{R}^{mn} + A \left(  r\right)  \tilde{e}^{m}%
\tilde{e}^{n}=0~,\label{BaseEqn}%
\end{equation}
where
\begin{equation}
A \left(  r\right)  := 60\alpha_{0}r^{4}+\frac{\alpha_{2} r^{2}}{f}\left(
3\left(  g^{4}\right)  ^{\prime}f^{\prime}+4g^{4}f^{\prime\prime}\right) \\
-3\alpha_{1} r^{2}\left(  2\left(  g^{2}r\right)  ^{\prime}+4g^{2}%
\frac{f^{\prime}}{f}r+\left(  g^{2}\right)  ^{\prime}\frac{f^{\prime}}{f}%
r^{2}+2g^{2}\frac{f^{\prime\prime}}{f}r^{2}\right)
\end{equation}
and
\begin{equation}
B := 2r^{2}\left[  3\alpha_{1}-\alpha_{2}\left(  \left(  g^{2}\right)
^{\prime}\frac{f^{\prime}}{f}+2g^{2}\frac{f^{\prime\prime}}{f}\right)
\right]
\end{equation}

In what follows we solve the field equations (\ref{eps1sinfac}) and
(\ref{BaseEqn}), starting from the generic case~(i), i.e., base manifolds with
a constant Ricci scalar $\tilde{R}=6\gamma$, and $g^{2}$ given by
(\ref{gcuagenconlambda}).

\bigskip

\noindent$\circ$ \emph{Radial and angular equations, Generic case (i):} The
radial field equation $\mathcal{E}_{1}=0$ allows to find the explicit form of
the function $f^{2}\left(  r\right)  $, whereas the components of the field
equations along the base manifold restricts its geometry to be of constant
curvature. This is seen as follows:\newline Since in case (i) the base
manifold has $\tilde{R}=6\gamma$, where $\gamma$ is a constant,
Eq.(\ref{eps1sinfac}) reads
\begin{equation}
\left(  B_{0}\left(  r\right)  -B_{1}\left(  r\right)  \right)  \gamma+\left(
A_{0}\left(  r\right)  -A_{1}\left(  r\right)  \right)  =0~,\label{propE1}%
\end{equation}
its only solution being $f^{2}=Cg^{2}$, where the constant $C$ can be absorbed
into a time rescaling. Thus, in the generic case (i), the solution to the
field equations $\mathcal{E}_{0}=\mathcal{E}_{1}=0$ for the ansatz
(\ref{Ansatz}) is $f^{2}=g^{2}$ given in (\ref{gcuagenconlambda})

The angular equations (\ref{BaseEqn}) imply
\begin{equation}
A\left(  r\right)  =-\lambda B\left(  r\right)  \ ,\label{ampropbm}%
\end{equation}
for some constant $\lambda$, and then (\ref{BaseEqn}) is equivalent to
\begin{equation}%
\begin{cases}
\label{ang}B(r)\left(  \tilde{R}^{mn}-\lambda\tilde{e}^{m}\tilde{e}%
^{n}\right)  =0,\\
A\left(  r\right)  =-\lambda B\left(  r\right)
\end{cases}
\end{equation}
Since $B(r)\neq0$ for $f^{2}=g^{2}$ given by (\ref{gcuagenconlambda}), the
base manifold must necessarily be of constant curvature, i.e., the metric of
$\Sigma_{3}$ satisfies $\tilde{R}^{mn}=\lambda\tilde{e}^{m}\tilde{e}^{n}$,
and, since $\tilde{R}=6\gamma$, it must be $\lambda=\gamma$. This takes care
of the first of equations (\ref{ang}). The second one adds nothing new since
\begin{equation}
A\left(  r\right)  +\gamma B\left(  r\right)  =0,\label{ang2}%
\end{equation}
is trivially satisfied because for $f=g$,
\begin{equation}
r^{-2}[A\left(  r\right)  +\gamma B\left(  r\right)  ]=r^{-1}[A_{0}\left(
r\right)  +\gamma B_{0}\left(  r\right)  ]^{\prime}\ ,
\end{equation}
and $g$ satisfies (\ref{apropbconst}). This concludes the classification of
case (i).\newline

\bigskip

\noindent$\circ$ \emph{Radial and angular equations, Special case (ii):} From
the constraint equation $\mathcal{E}_{0}=0$, one knows that in this case, the
Gauss-Bonnet coefficient is fixed as in Eq. (\ref{tuneo}), and the metric
function $g^{2}$ is given by Eq. (\ref{gbhspecial}).

The radial field equation (\ref{eps1sinfac}) now reads%
\begin{equation}
\left(  \left[  \mu-\sigma r^{2}\right]  \frac{f^{\prime}}{f}+\sigma r\right)
\left(  \tilde{R}+6\mu\right)  =0~,\label{eps1barfactorized}%
\end{equation}
which is solved either by

\bigskip

(ii.a) Having the first factor in (\ref{eps1barfactorized}) vanish, or by

(ii.b) Requiring the Ricci scalar of $\Sigma_{3}$ to be $\tilde{R}=-6\mu$.

\noindent After a time re-scaling, the solution in case (ii.a), is
$f^{2}=g^{2}$, (given in Eq. (\ref{gbhspecial})).

No restriction on $\Sigma_{3}$ is imposed in this case.

\noindent Case (ii.b), on the other hand, is solved by requiring $\tilde
{R}=-6\mu$, so that the scalar curvature of the base manifold is related to
the constant of integration in (\ref{gbhspecial}). Note that, in this case,
the metric function $f^{2}$ is left undetermined by the system $\mathcal{E}%
_{0} = \mathcal{E}_{1}=0$.

The remaining fields equations, $\mathcal{E}_{m}=0$, can be written as%
\begin{equation}
\left(  \sigma-\sigma r\frac{f^{\prime}}{f}-\left(  \sigma r^{2}-\mu\right)
\frac{f^{\prime\prime}}{f}\right)  \left(  \tilde{R}^{mn}+\mu\tilde{e}%
^{m}\tilde{e}^{n}\right)  =0~.\label{epsibarfactorized}%
\end{equation}

For case (ii.a), the first factor of Eq. (\ref{epsibarfactorized}) vanishes,
and the geometry of base manifold $\Sigma_{3}$ is left unrestricted. We have a
solution of the full set of field equations of the special theories
(\ref{tuneo}) given by (\ref{Ansatz}) with $f^{2}=g^{2}$ of Eq.
(\ref{gbhspecial}), and an arbitrary base manifold $\Sigma_{3}$.\newline In
case (ii.b), Eq.(\ref{epsibarfactorized}) can be solved in two different ways:

\bigskip

\noindent(ii.b1) Choosing $f$ such that the first factor vanishes.
\newline(ii.b2) Requiring the base manifold to be of constant curvature $-\mu
$, i.e., $\tilde{R}^{mn}= - \mu\tilde{e}^{m}\tilde{e}^{n}$.

\bigskip

\noindent Case (ii.b2) leaves the redshift function $f^{2}$ completely
undetermined.\newline Case (ii.b1) opens new interesting possibilities. The
vanishing of the first factor of Eq. (\ref{epsibarfactorized}) gives a
differential equation for the redshift function, whose general solution, after
a time rescaling, reads
\begin{equation}
f^{2}(r)=\left\{
\begin{array}
[c]{ccc}%
\left(  \sqrt{\sigma}r+a\sqrt{\sigma r^{2}-\mu}\right)  ^{2} & : & \mu\neq0\\
\left(  a\sqrt{\sigma}r+\frac{1}{\sqrt{\sigma}r}\right)  ^{2} & : & \mu=0
\end{array}
\right.  ~,
\end{equation}
where $a$ is an integration constant. $\Sigma_{3}$ is not a constant curvature
manifold, although it has constant Ricci scalar $\tilde{R}=-6\mu$. Note that
we do not loose generality if we set $-\mu$ equal to $\gamma=\pm1,0 $.

For $\gamma\neq0$ there are three distinct cases, namely $|a|>1$, $|a|<1$ or
$|a|=1$, with substantially different qualitative features. It is simple to
show that, for negative cosmological constant $\left(  \sigma>0\right)  $, the
solution with $\gamma=-1$ and $|a|<1$ corresponds to the wormhole in vacuum
found in \cite{DOTwormhole}, whereas that with $|a|=1$ corresponds to a brand
new wormhole in vacuum (See Section III).\newline On the other hand, if
$\gamma=0$ (base manifold with vanishing Ricci scalar), for negative
cosmological constant and nonnegative $a$, the metric (\ref{Ansatz}) describes
a spacetime that looks like a \textquotedblleft spacetime horn". We will see
in the next section that if the base manifold is not locally flat, there is a
timelike naked singularity. Yet, the mass of the solution vanishes and the
Euclidean continuation has a finite action (See Section \ref{euc}).

\bigskip

This concludes our classification of solutions. Since case (i) has been
extensively discussed in the literature, we devote the following sections to a
discussion of the novel solutions (ii)a and (ii)b1/b2.

\section{Geometrically well behaved solutions: Black holes, wormholes and
spacetime horns}

\label{geom}

In this Section we study the solutions for the special case found above.

One can see that, when they describe black holes and wormholes, as $r$ goes to
infinity the spacetime metric approaches that of a spacetime of constant
curvature $-\sigma$, with different kinds of base manifolds. This is also the
case for spacetime horns, provided $a\neq0$ (See Sec. III. B). It is simple to
verify by inspection that for $\sigma\leq0$, the solutions within the special
case are geometrically ill-behaved in general. Hence, hereafter we restrict
our considerations to the case $l^{2}:=\sigma^{-1}>0$, where $l$ is the
anti-de Sitter (AdS) radius.

\subsection{\textit{Case (ii.a): Black holes}}

According to the classification presented in the previous section, fixing an
arbitrary base manifold $\Sigma_{3}$, the metric
\begin{equation}
ds^{2}=-\left(  \frac{r^{2}}{l^{2}}-\mu\right)  dt^{2}+\frac{dr^{2}}{\left(
\frac{r^{2}}{l^{2}}-\mu\right)  }+r^{2}d\Sigma_{3}^{2}~.\label{BH-metric}%
\end{equation}
solves the full set of Einstein Gauss Bonnet equations for the special
theories (\ref{tuneo}). The integration constant $\mu$ is related to the mass,
which is explicitly computed from a surface integral in Section \ref{mass}.
For $\mu>0$, the metric (\ref{BH-metric}) describes a black hole whose horizon
is located at $r=r_{+}:=\sqrt{\mu}\;l$. Requiring the Euclidean continuation
to be smooth, the black hole temperature can be obtained from the Euclidean
time period, which is given by
\begin{equation}
\beta=\frac{1}{T}=\frac{2\pi l^{2}}{r_{+}}\ .\label{Temperature}%
\end{equation}
For later purposes it is useful to express the Euclidean black hole solution
in terms of the proper radial distance $\rho$ (in units of $l$), given by%
\[
r=r_{+}\cosh(\rho)\ ,
\]
with $0\leq\rho<\infty$, so that the Euclidean metric reads%
\begin{equation}
ds^{2}=\frac{r_{+}^{2}}{l^{2}}\sinh^{2}(\rho)d\tau^{2}+l^{2}d\rho^{2}%
+r_{+}^{2}\cosh^{2}(\rho)d\Sigma_{3}^{2}~.\label{Euclidean BH}%
\end{equation}
The thermodynamics of these kind of black holes turns out to be very sensitive
to the geometry of the base manifold, this is briefly discussed in Section
\ref{euc}.

\subsection{\textit{Case (ii.b): Wormholes and spacetime horns}}

In this case the base manifold possesses a constant Ricci scalar $\tilde
{R}=6\gamma$, with $\gamma$ normalized to $\pm1$ or $0$.

Let us first consider the case for which the base manifold $\Sigma_{3}$ has
nonvanishing Ricci scalar, i.e., $\gamma\neq0$. By virtue of Eqs.
(\ref{fcuaworm}), and (\ref{gcuaesp}) the spacetime metric (\ref{Ansatz})
reads%
\begin{equation}
ds^{2}=-\left(  \frac{r}{l}+a\sqrt{\frac{r^{2}}{l^{2}}+\gamma}\right)
^{2}dt^{2}+\frac{dr^{2}}{\frac{r^{2}}{l^{2}}+\gamma}+r^{2}d\Sigma_{3}%
^{2}\ ,\label{wormholerre}%
\end{equation}
where $a$ is an integration constant and $l>0$. The Ricci scalar of
(\ref{wormholerre}) is given by%
\begin{equation}
R=-\frac{20}{l^{2}}-\frac{6\gamma}{l}\left[  r\left(  \frac{r}{l}+a\sqrt
{\frac{r^{2}}{l^{2}}+\gamma} \;\;\right)  \right]  ^{-1}\ ,\label{Rserre}%
\end{equation}
which generically diverges at $r=0$ and at any point satisfying $r/a<0$ and
\begin{equation}
r_{s}^{2}= l^{2} \; \; \frac{ \gamma\; a^{2}}{1-a^{2}}\ .\label{rs}%
\end{equation}

In the case $\gamma=1$ the metric possesses a timelike naked singularity at
$r=0$, and if $-1<a<0$, an additional timelike naked singularity at
$r^{2}=r_{s}^{2}$. Due to this ill geometrical behavior, we no longer consider
the spacetime (\ref{wormholerre}) for the case $\gamma=1$.

\bigskip

$\circ$ \emph{Wormholes: }The case $\gamma=-1$ is much more interesting. The
region $r<l$ must be excised since the metric (\ref{wormholerre}) becomes
complex within this range, and the Schwarzschild-like coordinates in
(\ref{wormholerre}) fail at $r=l$. Introducing the proper radial distance
$\rho$, given by%
\[
r=l\cosh\left(  \rho\right)  \ .
\]
allows to extend the manifold beyond $r=l$ ($\rho>0$) to a geodesically
complete manifold by letting $-\infty<\rho<\infty$. For $a^{2}<1$ the
resulting metric for this geodesically complete manifold reads%
\begin{equation}
ds^{2}=l^{2}\left[  -\cosh^{2}\left(  \rho-\rho_{0}\right)  dt^{2}+d\rho
^{2}+\cosh^{2}\left(  \rho\right)  d\Sigma_{3}^{2}\right]  \ ,\label{DOTworm}%
\end{equation}
where $\rho_{0}:=-\tanh^{-1}(a)$, and the time coordinate has been rescaled.
Note that since (\ref{wormholerre}) is invariant under $(r,a)\rightarrow
(-r,-a)$, the $\rho>0$ piece of (\ref{DOTworm}) is isometric to
(\ref{wormholerre}) whereas the $\rho<0$ portion is isometric to the metric
obtained by replacing $a\rightarrow-a$ in (\ref{wormholerre}). In other words,
(\ref{DOTworm}) matches the region $r\geq l$ of the metric (\ref{wormholerre})
with a given value of $a$, with the region $r\geq l$ of the same metric but
reversing the sign of $a$. The singularity at $r^{2}=r_{s}^{2}$ in Eq.
(\ref{rs}) is not present since $a^{2}\leq1$, and that at $r=0$ is also absent
since $r\geq l>0$ at all points.

For $a^{2}=1$ we obtain another wormhole in vacuum, by using again the proper
distance $\rho$ defined above:%

\begin{equation}
ds^{2}=l^{2}\left[  -e^{2\rho}dt^{2}+d\rho^{2}+\cosh^{2}\left(  \rho\right)
d\Sigma_{3}^{2}\right]  \ .\label{worm2}%
\end{equation}

In these coordinates it is manifest that the metrics (\ref{DOTworm}) and
(\ref{worm2}) describe wormholes, both possessing a throat located at $\rho
=0$. No energy conditions are violated by these solutions, since in both
cases, the whole spacetime is devoid of any kind of stress-energy
tensor.\bigskip

\noindent The spacetime described by Eq. (\ref{DOTworm}) is the static
wormhole solution found in \cite{DOTwormhole}. This metric connects two
asymptotically locally AdS regions, and gravity pulls towards a fixed
hypersurface located at $\rho=\rho_{0}$ being parallel to the neck. This is
revisited in the next subsection.

\bigskip

The metric (\ref{worm2}) describes a brand new wormhole. Its Riemann tensor is
given by%

\begin{align}
\label{curv1}R_{\ \ t\rho}^{t\rho}  &  =-\frac{1}{l^{2}}\ ,\ R_{\ \ \rho
j}^{\rho i}=-\frac{1}{l^{2}}\delta_{j}^{i}\ ,\ R_{\ \ tj}^{ti}=-\frac{1}%
{l^{2}}\tanh\left(  \rho\right)  \delta_{j}^{i}\ ,\nonumber\\
R_{\ \ kl}^{ij}  &  =\frac{1}{l^{2}}\frac{\tilde{R}_{\ \ kl}^{ij}}{\cosh
^{2}\left(  \rho\right)  }-\frac{1}{l^{2}}\tanh^{2}\left(  \rho\right)
\left(  \delta_{k}^{i}\delta_{l}^{j}-\delta_{l}^{i}\delta_{k}^{j}\right)  \ ,
\end{align}
where latin indices run along the base manifold. At the asymptotic regions
$\rho\rightarrow\pm\infty$ the curvature components approach
\begin{align}
\label{Rare-curvatures}R_{\ \ t\rho}^{t\rho}  &  =-\frac{1}{l^{2}%
}\ ,\ R_{\ \ \rho j}^{\rho i}=-\frac{1}{l^{2}}\delta_{j}^{i}\ ,\ R_{\ \ tj}%
^{ti}\simeq\mp\frac{1}{l^{2}}\delta_{j}^{i}\ ,\nonumber\\
R_{\ \ kl}^{ij}  &  \simeq-\frac{1}{l^{2}}\left(  \delta_{k}^{i}\delta_{l}%
^{j}-\delta_{l}^{i}\delta_{k}^{j}\right)  \ ,
\end{align}

This makes clear that the wormhole (\ref{worm2})\ connects an asymptotically
locally AdS spacetime (at $\rho\rightarrow\infty$) with another nontrivial
smooth spacetime at the other asymptotic region ($\rho\rightarrow-\infty$).
Note that although the metric looks singular at $\rho\rightarrow-\infty$, the
geometry is well behaved at this asymptotic region. This is seen by noting
that the basic scalar invariants can be written in terms of contractions of
the Riemann tensor with the index position as in (\ref{curv1}), whose
components have well defined limits (given in (\ref{Rare-curvatures})), and
$g^{\alpha}{}_{\beta} = \delta^{\alpha}{}_{\beta}$. Thus, the invariants
cannot diverge. As an example, the limits of some invariants are
\begin{equation}
\lim_{\rho\rightarrow-\infty} R^{\alpha\beta}{}_{\alpha\beta} = -\frac
{8}{l^{2}}\ ,\ \lim_{\rho\rightarrow-\infty}R^{\alpha\beta}{}_{\gamma\delta
}R^{\gamma\delta}{}_{\alpha\beta}=\frac{40}{l^{4}}\; , \; \lim_{\rho
\rightarrow-\infty}C^{\alpha\beta}{}_{\gamma\delta}C^{\gamma\delta}{}%
_{\alpha\beta}=\frac{8}{l^{4}}\
\end{equation}
where $C^{\alpha\beta}{}_{\gamma\delta}$ is the Weyl tensor. \newline We have
also computed some differential invariants and found they are all well behaved
as $\rho\rightarrow-\infty$.

Some features about the geodesics in these vacuum wormholes are discussed in
the next subsection, their regularized Euclidean actions and their masses are
evaluated in Sections \ref{mass} and \ref{euc}, respectively.

\bigskip

\noindent$\circ$ \emph{Spacetime horns: }Let us consider now the case when the
base manifold $\Sigma_{3}$ has vanishing Ricci scalar, i.e., $\tilde{R}=0 $.

In this case the metric (\ref{Ansatz}) reduces to%
\begin{equation}
ds^{2}=-\left(  a\frac{r}{l}+\frac{l}{r}\right)  ^{2}dt^{2}+l^{2}\frac{dr^{2}%
}{r^{2}}+r^{2}d\Sigma_{3}^{2}\ ,\label{g-horn}%
\end{equation}
where $a$ is an integration constant. The Ricci scalar of this spacetime reads%
\begin{equation}
R=-\frac{4}{l^{2}}\left(  \frac{5ar^{2}+l^{2}}{l^{2}+ar^{2}}\right)  \ .
\end{equation}
The timelike naked singularity at $r_{s}^{2}=-\frac{l^{2}}{a}$ can be removed
requiring $a\geq0$; however this condition is not strong enough to ensure that
the spacetime is free of singularities. Indeed the Kretschmann scalar is given
by%
\begin{equation}
K:=R_{\lambda\rho}{}^{\mu\nu}R_{\mu\nu}{}^{\lambda\rho}=\frac{\tilde{R}_{kl}%
{}^{ij}\tilde{R}_{ij}{}^{kl}}{r^{4}}+\frac{8\left(  5r^{4}a^{2}+4l^{2}%
r^{2}a+5l^{4}\right)  }{l^{4}\left(  ar^{2}+l^{2}\right)  ^{2}}%
\ ,\label{kretchhorned}%
\end{equation}
where $\tilde{R}_{kl}{}^{ij}\tilde{R}_{ij}^{\ \ kl}$ is the Kretchmann scalar
of the Euclidean base manifold $\Sigma_{3}$. Hence, for a generic base
manifold with vanishing Ricci scalar, the metric possesses a timelike naked
singularity at $r=0$, unless the Kretchmann scalar of the base manifold
vanishes. Since the base manifold is Euclidean, the vanishing of its
Kretchmann scalar implies that it is locally flat. This drives us out of
(ii.b1) to the degenerate case (ii.b2), for which the $g_{tt}$ component of
the metric is not fixed by the field equations, for this reason we will not
consider the locally flat case.

If the base manifold is not locally flat, at the origin the Ricci scalar goes
to a constant and the Kretschmann scalar diverges as $r^{-4}$. Therefore, the
singularity at the origin is smoother than that of a conifold \cite{Candelas},
whose Ricci scalar diverges as $r^{-2}$, and it is also smoother than that of
the five-dimensional Schwarzschild metric with negative mass, that possesses a
timelike naked singularity at the origin with a Kretschmann scalar diverging
as $r^{-8}$. In spite of this divergency, the regularized Euclidean action and
the mass are finite for this solution, as we show in Sections \ref{euc} and
\ref{mass}. In this sense this singularity is as tractable as that of a vortex.

\bigskip

In the case $a>0$ we are interested in, we introduce $a=:e^{-2\rho_{0}}$ and a
time rescaling; then the metric (\ref{g-horn}) expressed in terms of the
proper radial distance $r=le^{\rho}$ is
\begin{equation}
ds^{2}=l^{2}\left[  -\cosh^{2}\left(  \rho-\rho_{0}\right)  dt^{2}+d\rho
^{2}+e^{2\rho}d\Sigma_{3}^{2}\right]  .\label{horn1}%
\end{equation}
This spacetime possesses a single asymptotic region at $\rho\rightarrow
+\infty$ where it approaches AdS spacetime, but with a base manifold different
from $S^{3}$. Note that as the warp factor of the base manifold goes to zero
exponentially as $\rho\rightarrow-\infty$, it actually looks like a
\textquotedblleft spacetime horn".

\bigskip

For $a=0$, the metric (\ref{g-horn}) can also be brought into the form of a
spacetime horn,%
\begin{equation}
ds^{2}=l^{2}\left[  -e^{-2\rho}dt^{2}+d\rho^{2}+e^{2\rho}d\Sigma_{3}%
^{2}\right]  \ \label{horn2}%
\end{equation}
which also possesses a single asymptotic region at $\rho\rightarrow+\infty$,
which agrees with the asymptotic form of the new wormhole (\ref{worm2}) as
$\rho\rightarrow-\infty$.\newline The asymptotic form of the Riemann tensor is
not that of a constant curvature manifold, and can then be obtained from the
$\rho\rightarrow-\infty$ limit in (\ref{Rare-curvatures}).

The regularized Euclidean action and Mass of these spacetime horns are
evaluated in Sections \ref{euc} and \ref{mass}. Geodesics are discussed in the
next subsection.

\subsection{Geodesics around wormholes and spacetime horns}

The class of metrics that describe the wormholes and spacetime horns is of the
form%
\begin{equation}
ds^{2}=-A^{2}\left(  \rho\right)  dt^{2}+l^{2}d\rho^{2}+C^{2}\left(
\rho\right)  d\Sigma^{2}\ ,\label{genericmetricrecu}%
\end{equation}
where the functions $A\left(  \rho\right)  $ and $C\left(  \rho\right)  $ can
be obtained from Eqs. (\ref{DOTworm}) and (\ref{worm2}) for wormholes, and
from Eqs.(\ref{horn1}) and (\ref{horn2}) for spacetime horns.

\subsubsection{Radial geodesics}

Let us begin with a brief analysis of radial geodesics for the wormholes and
spacetime horns. The radial geodesics are described by the following equations%
\begin{align}
\dot{t}-\frac{E}{A^{2}}  &  =0\text{\ },\label{tdot}\\
l^{2}\dot{\rho}^{2}-\frac{E^{2}}{A^{2}}+b  &  =0\ ,\label{rhodot}%
\end{align}
where dot stands for derivatives with respect to the proper time, the velocity
is normalized as $u_{\mu}u^{\mu}=-b$, and the integration constant $E$
corresponds to the energy. As one expects, Eq. (\ref{rhodot}) tells that
gravity is pulling towards the fixed hypersurface defined by $\rho=\rho_{0}$,
where $\rho_{0}$ is a minimum of $A^{2}\left(  \rho\right)  $.

\bigskip

$\circ$ \emph{Wormholes: } From (\ref{DOTworm}) we have $A^{2}(\rho
)=l^{2}\cosh^{2}(\rho-\rho_{0})$, then the equations for radial geodesics
(\ref{tdot}) and (\ref{rhodot}) reduce to
\begin{align}
\dot{\rho}^{2}-\frac{E^{2}}{l^{4}\;\cosh^{2}\left(  \rho-\rho_{0}\right)  }
&  =-\frac{b}{l^{2}}\ ,\\
\dot{t}-\frac{E}{l^{2}\;\cosh^{2}\left(  \rho-\rho_{0}\right)  }  &  =0\ .
\end{align}

These equation immediately tell us that \cite{DOTwormhole}: The $\rho$
coordinate of a radial geodesic behaves as a classical particle in a
P\"{o}schl-Teller potential; timelike geodesics are confined, they oscillate
around the hypersurface $\rho=\rho_{0}$. An observer sitting at $\rho=\rho
_{0}$ lives in a timelike geodesic (here $d\tau/dt=l,$ $\tau$ the proper time
of this static observer); radial null geodesics connect both asymptotic
regions (i.e., $\rho=-\infty$ with $\rho=+\infty$) in a finite $t$ span
$\Delta t=\pi$, which does not depend on $\rho_{0}$ (the static observer at
$\rho=\rho_{0}$ says that this occurred in a proper time $\Delta\tau=\pi l$).
These observations give a meaning to $\rho_{0}$: gravity is pulling towards
the fixed hypersurface defined by $\rho=\rho_{0}$, which is parallel to the
neck at $\rho=0$, and therefore $\rho_{0}$ is a modulus parameterizing the
proper distance from this hypersurface to the neck.

\bigskip

The geodesic structure of the new wormhole (\ref{worm2}) is quite different
from the previous one. In this case, the radial geodesic Eqs. (\ref{tdot}) and
(\ref{rhodot}) read%
\begin{align}
\dot{\rho}^{2}-\frac{e^{-2\rho}E^{2}}{l^{4}}  &  =-\frac{b}{l^{2}}\ ,\\
l^{2}\;\dot{t}-e^{-2\rho}E  &  =0\ .
\end{align}
As expected, the behavior of the geodesics at $\rho\rightarrow+\infty$ is like
in an AdS spacetime. Moreover, since gravity pulls towards the asymptotic
region $\rho\rightarrow-\infty$, radial timelike geodesics always have a
turning point and they are doomed to approach to $\rho\rightarrow-\infty$ in
the future. Note that the proper time that a timelike geodesic takes to reach
the asymptotic region at $\rho=-\infty$, starting from $\rho=\rho_{f}$ is
finite and given by
\begin{equation}
\Delta\tau=\int_{\rho\ =\ -\infty}^{\rho\ =\ \rho_{f}}\frac{l^{2}d\rho}%
{\sqrt{E^{2}e^{-2\rho}-l^{2}}}=\frac{\pi l}{2}-l\tan^{-1}\left(  \sqrt
{\frac{E^{2}}{l^{2}}e^{-2\rho_{f}}-1}\right)  <\infty\ .\label{fpt}%
\end{equation}
It is easy to check that null radial geodesics can also reach the asymptotic
region at $\rho=-\infty$ in a finite affine parameter. This, together with the
fact that spacetime is regular at this boundary, seems to suggest that it
could be analytically continued through this surface. However, since the warp
factor of the base manifold blows up at $\rho=-\infty$, this null hypersurface
should be regarded as a spacetime boundary.

\bigskip

$\circ$ \emph{Spacetime horns}: For the spacetime horn (\ref{horn1}), the
($\rho,t$) piece of the metric agrees with that of the wormhole (\ref{DOTworm}%
). Hence, the structure of radial geodesics in both cases is the same, with
gravity pulling towards the $\rho=\rho_{0}$ surface. Timelike geodesics again
have a turning point, which, in this case, prevents the geodesics from hitting
the singularity at $\rho=-\infty$.

\bigskip

In the case of the spacetime horn (\ref{horn2}) (compare to (\ref{worm2})),
gravity becomes a repulsive force pointing from the singularity at
$\rho\rightarrow-\infty$, towards the asymptotic region at $\rho
\rightarrow+\infty$. Therefore timelike radial geodesics are doomed to end up
at the asymptotic region in a finite proper time (see (\ref{fpt}))

\subsubsection{Gravitational vs. centrifugal forces}

In this Section we discuss an interesting effect that occurs for geodesics
with nonzero angular momentum. One can see that for the generic class of
spacetimes (\ref{genericmetricrecu}), which includes wormholes and spacetime
horns, there is a region where the gravitational and centrifugal effective
forces point in the same direction. These are expulsive regions that have a
single turning point for any value of the conserved energy, and within which
bounded geodesics cannot exist.

The class of metrics we consider are (\ref{genericmetricrecu}) with the
further restriction that the base manifold $\Sigma_{3}$ have a Killing vector
$\xi$. Choosing adapted coordinates $y=(x^{1},x^{2},\phi)$ such that
$\xi=\partial/\partial\phi$, the base manifold metric is $d\Sigma_{3}%
^{2}=\tilde{g}_{ij}(x)dy^{i}dy^{j}$ and the spacetime geodesics with $x$ fixed
are described by the following equations%
\begin{align}
\dot{t}  &  =\frac{E}{A^{2}}\text{\ ;\ }\dot{\phi}=\frac{L}{C^{2}}\nonumber\\
l^{2}\dot{\rho}^{2}  &  =-b+\frac{E^{2}}{A^{2}}-\frac{L^{2}}{C^{2}%
}.\label{radialdepencegeodesic}%
\end{align}
Here we have used the fact that, if $u^{a}$ is the geodesic tangent vector,
then $\xi^{a}u_{a}=\mathcal{L}$ is conserved, and $\dot{\phi}=\mathcal{L}%
/(C^{2}\tilde{g}_{\phi\phi}(x))=:L/C^{2}$. If $\xi$ is a $U(1)$ Killing vector
then $\mathcal{L}$ is a conserved angular momentum. Examples are not hard to
construct, for spacetime horns, what we need is a base manifold with zero
Ricci scalar and a $U(1)$ Killing field. For wormholes, we need a nonflat
3-manifold with $\tilde{R}=-6$ and a $U(1)$ isometry, an example being
$S^{1}\times H_{2}/\Gamma$, where $\Gamma$ is a freely acting discrete
subgroup of $O(2,1)$, and the metric locally given by:
\begin{equation}
d\Sigma_{3}^{2}=\frac{1}{3}\left(  dx_{1}{}^{2}+\sinh^{2}(x_{1})\;dx_{2}{}%
^{2}\right)  +d\phi^{2}\ .
\end{equation}

The motion along the radial coordinate in proper time is like that of a
classical particle in an effective potential given by the r.h.s. of Eq.
(\ref{radialdepencegeodesic}). This effective potential, has a minimum at
$\rho=\bar{\rho}$ only if the following condition is fulfilled%
\begin{equation}
\frac{A^{\prime}\left(  \bar{\rho}\right)  }{A\left(  \bar{\rho}\right)  ^{3}%
}E^{2}=\frac{C^{\prime}\left(  \bar{\rho}\right)  }{C\left(  \bar{\rho
}\right)  ^{3}}L^{2}.\label{minpot}%
\end{equation}
This expresses the fact that the gravitational effective force is canceled by
the centrifugal force if the orbit sits at $\rho=\bar{\rho}$. The class of
spacetimes under consideration have regions $\mathcal{U}$ where the sign of
$A^{-3}A^{\prime}$ is opposite to that of $C^{-3}C^{\prime}$, i.e., the
effective gravitational and centrifugal forces point in the same direction.
Within these regions, there is at most a single turning point, and
consequently bounded orbits cannot exist.

In the case of a wormhole (\ref{DOTworm}), Eq. (\ref{minpot}) reads
\begin{equation}
\frac{E^{2}\tanh\left(  \bar{\rho}-\rho_{0}\right)  }{\cosh^{2}\left(
\bar{\rho}-\rho_{0}\right)  }=\frac{L^{2}\tanh\bar{\rho}}{\cosh^{2}\bar{\rho}%
}\ .\label{restrictsgeo}%
\end{equation}
The centrifugal force reverses its sign at the neck at $\rho=0$, the Newtonian
force does it at $\rho=\rho_{0}$, both forces being aligned for $\rho$ between
zero and $\rho_{0}$. The expulsive region $\mathcal{U}$ is nontrivial whenever
$\rho_{0}\neq0$. This situation is depicted in Fig. 1a.

In the case of the new wormhole solution (\ref{worm2}) the region
$\mathcal{U}$ is defined $\rho\leq0$ (See Fig. 1b), and for the spacetime horn
(\ref{horn1}) the region $\mathcal{U}$ is given by $\rho\leq\rho_{0}$ (Fig.
1c). Finally, for the spacetime horn (\ref{horn2}) the region $\mathcal{U}$ is
the entire spacetime, there are no bounded geodesics.

\begin{figure}[ptb]
\includegraphics[scale=0.60,angle=-90]{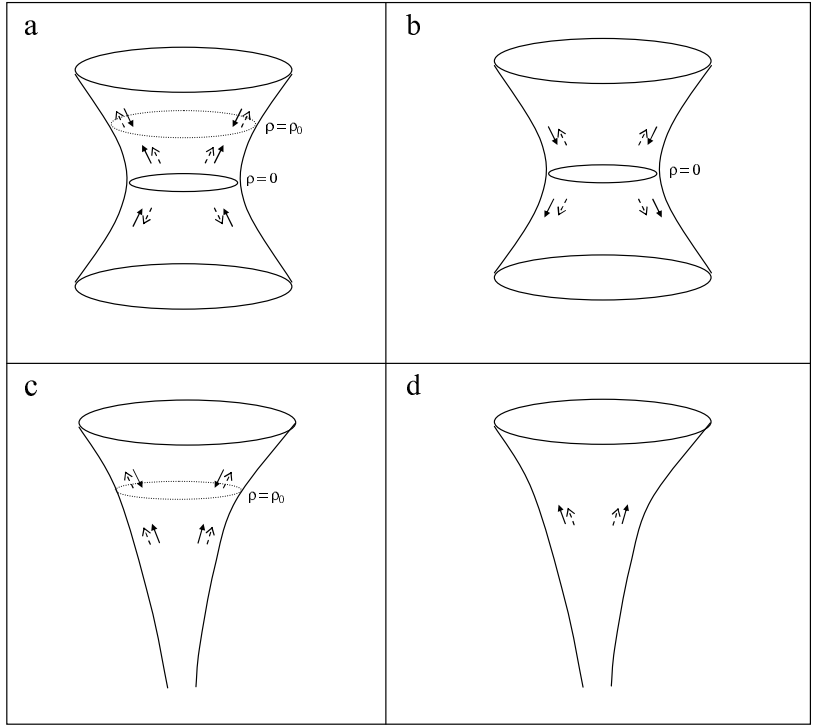}
\caption{Gravitational vs. centrifugal forces for wormholes and spacetime
horns. In this diagram, black and dashed arrows represent effective
gravitational and centrifugal forces, respectively. Figures a and b,
correspond to the wormholes (\ref{DOTworm}) and (\ref{worm2}), while figures c
and d represent the spacetime horns (\ref{horn1}) and (\ref{horn2}),
respectively.}%
\label{fig:epsart}%
\end{figure}

\section{Regularized Euclidean action}

\label{euc}

Here it is shown that the geometrically well-behaved solutions discussed in
the previous Section have finite Euclidean action, which reduces to the free
energy in the case of black holes, and vanishes for the other solutions.

The action (\ref{action}) in the case of special choice of coefficients can be
written as%
\begin{equation}
I_{5}=\kappa\!\int_{M}\!\epsilon_{abcde}\left(  \!R^{ab}R^{cd}\!+\!\frac
{2}{3l^{2}}R^{ab}e^{c}e^{d}\!+\!\frac{1}{5l^{4}}e^{a}e^{b}e^{c}e^{d}\!\right)
\!e^{e}\ ,\label{Ibulk}%
\end{equation}
and it has been shown that it can be regularized by adding a suitable boundary
term in a background independent way, which depends only on the extrinsic
curvature and the geometry at the boundary \cite{MOTZ}. The total action then
reads
\begin{equation}
I_{T}=I_{5}-B_{4}\ ,\label{It}%
\end{equation}
where the boundary term is given by%
\begin{equation}
B_{4}\!=\!\kappa\!\int_{\partial M}\!\epsilon_{abcde}\theta^{ab}e^{c}\left(
\!R^{de}-\frac{1}{2}\theta_{\ f}^{d}\theta^{fe}+\frac{1}{6l^{2}}e^{d}%
e^{e}\!\right)  \ ,\label{BoundaryTerm5}%
\end{equation}
and $\theta^{ab}$ is the second fundamental form. The total action (\ref{It})
attains an extremum for solutions of the field equations provided%
\begin{equation}
\delta I_{T}=\kappa\int_{\partial M}\epsilon_{abcde}\left(  \delta\theta
^{ab}e^{c}-\theta^{ab}\delta e^{c}\right)  \left(  \bar{R}^{de}-\frac{1}%
{2}\theta_{\ f}^{d}\theta^{fe}-\frac{1}{2l^{2}}e^{d}e^{e}\right)
=0\ ,\label{boundarycondition}%
\end{equation}
where $\bar{R}^{ab}:=R^{ab}+\frac{1}{l^{2}}e^{a}e^{b}$. Therefore, the value
of the regularized Euclidean action makes sense for solutions which are
\textit{bona fide} extrema, i.e., for solutions such that condition
(\ref{boundarycondition}) is fulfilled.

The Euclidean continuation of the class of spacetimes described in Section
III, including black holes, wormholes and spacetime horns, is described by
metrics of the form%
\begin{equation}
ds^{2}=A^{2}\left(  \rho\right)  d\tau^{2}+l^{2}d\rho^{2}+C^{2}\left(
\rho\right)  d\Sigma_{3}^{2}\ ,\label{genericabmetric}%
\end{equation}
where $0\leq\tau\leq\beta$ is the Euclidean time, and the functions $A$ and
$C$ correspond to the ones appearing in Eq. (\ref{Euclidean BH}) for the black
holes; Eqs. (\ref{DOTworm}) and (\ref{worm2}) for the wormholes, and in Eqs.
(\ref{horn1}) and (\ref{horn2}) for the spacetime horns.

Let us first check that these solutions are truly extrema of the total action
(\ref{It}).

\subsection{Geometrically well-behaved solutions as extrema of the regularized
action}

For the class of solutions under consideration, the curvature two-form
satisfies%
\begin{equation}
\bar{R}^{01}=\bar{R}^{1m}=0\ ,\label{curvcond}%
\end{equation}
and the condition (\ref{boundarycondition}) reduces to%
\begin{equation}
\delta I_{T}=\kappa\beta\left[  F\ \mathcal{I}_{3}+6\ G\ \mathcal{V}%
_{3}\right]  _{\partial\Sigma}\ ,\label{variationactionabc}%
\end{equation}
where $\beta\,$is the Euclidean time period, $\mathcal{V}_{3}$ is the volume
of the base manifold, and $\partial\Sigma$ is the boundary of the spatial
section. In Eq. (\ref{variationactionabc}) $\mathcal{I}_{3}$ is defined by
\begin{equation}
\mathcal{I}_{3}:=\int_{\Sigma_{3}}\sqrt{\tilde{g}}\tilde{R}\ d^{3}x\ ,
\end{equation}
and the functions $F$ and $G$ in (\ref{variationactionabc}) are given by%
\begin{align}
F  &  :=\frac{2}{l}\left[  A^{\prime}\delta C-A\delta C^{\prime}+C^{\prime
}\delta A-C\delta A^{\prime}\right]  \ ,\label{Fvariation}\\
G  &  :=\left[  A^{\prime}\left(  C^{2}-C^{\prime2}\right)  +2C^{\prime
}\left(  CA-C^{\prime}A^{\prime}\right)  \right]  \frac{\delta C}{l^{3}%
}\nonumber\\
&  -\left[  A\left(  C^{2}-C^{\prime2}\right)  +2C\left(  CA-C^{\prime
}A^{\prime}\right)  \right]  \frac{\delta C^{\prime}}{l^{3}}\label{Gvariation}%
\\
&  +C^{\prime}\left(  C^{2}-C^{\prime2}\right)  \frac{\delta A}{l^{3}%
}-C\left(  C^{2}-C^{\prime2}\right)  \frac{\delta A^{\prime}}{l^{3}%
}\ .\nonumber
\end{align}
Here we work in the minisuperspace approach, where the variation of the
functions $A$ and $C$ correspond to the variation of the integration
constants, and prime $\left(  ^{\prime}\right)  $ denotes derivative with
respect to $\rho$.

Now it is simple to evaluate the variation of the action
(\ref{variationactionabc}) explicitly for each case.

\bigskip

$\circ$ \emph{Black holes: }As explained in Section III, the Euclidean black
hole metric is given by%
\begin{equation}
ds^{2}=\frac{r_{+}^{2}}{l^{2}}\sinh^{2}(\rho)d\tau^{2}+l^{2}d\rho^{2}%
+r_{+}^{2}\cosh^{2}(\rho)d\Sigma_{3}^{2}\ ,\label{euclbh}%
\end{equation}
with $\beta=\frac{2\pi l^{2}}{r_{+}}$, and it has a single boundary which is
of the form $\partial M=S^{1}\times\Sigma_{3}$. In order to evaluate
(\ref{variationactionabc}) it is useful to introduce the regulator $\rho_{a}$,
such that $0\leq\rho\leq\rho_{a}$. It is easy to verify that the functions $F
$ and $G$ defined in (\ref{Fvariation}) and (\ref{Gvariation}) respectively,
satisfy%
\begin{equation}
F\left(  \rho_{a}\right)  =G\left(  \rho_{a}\right)  =0\ ,
\end{equation}
and hence, the boundary term (\ref{variationactionabc}) identically vanishes.
Note that it was not necessary to take the limit $\rho_{a}\rightarrow+\infty$.

\bigskip

$\circ$ \emph{Wormholes: }The Euclidean continuation of both wormhole
solutions in Eqs. (\ref{DOTworm}) and (\ref{worm2}) can be written as%
\begin{equation}
ds^{2}=l^{2}\left[  \left(  \cosh\rho+a\sinh\rho\right)  ^{2}d\tau^{2}%
+d\rho^{2}+\cosh^{2}\rho d\Sigma_{3}^{2}\right]  \ ,\label{Euclideanworms}%
\end{equation}
where the metrics (\ref{DOTworm}) and (\ref{worm2}) are recovered for
$a^{2}<1$ and $a^{2}=1$, respectively, and $\beta$ is arbitrary. In this
sense, the wormhole (\ref{worm2}) can be regarded as a sort of extremal case
of the wormhole (\ref{DOTworm}). In this case, since the boundary is of the
form $\partial\Sigma=\Sigma_{3}^{+}\cup\Sigma_{3}^{-}$ it is useful to
introduce the regulators $\rho_{\pm}$, such that $\rho_{-}\leq\rho\leq\rho
_{+}\ $. Using the fact that the base manifold has a negative constant Ricci
scalar given by $\tilde{R}=-6$, the variation of the action
(\ref{variationactionabc}) reduces to%
\begin{equation}
\delta I_{T}=6\kappa\beta l\ \delta a\ \left[  \mathcal{V}_{3}\right]
_{\rho_{-}}^{\rho_{+}}=0\ .
\end{equation}
Note that, as in the case for the black hole, the boundary term vanishes
regardless the position of the regulators $\rho_{-}$ and $\rho_{+}$.

\bigskip

$\circ$ \emph{Spacetime horns: }The Euclidean continuation of the spacetime
horns in Eqs. (\ref{horn1}) and (\ref{horn2}) can be written as%

\begin{equation}
ds^{2}=l^{2}\left[  \left(  ae^{\rho}+e^{-\rho}\right)  ^{2}d\tau^{2}%
+d\rho^{2}+e^{2\rho}d\Sigma_{3}^{2}\right]  \ ,\label{euclhorn1}%
\end{equation}
with an arbitrary time period $\beta$. The metrics (\ref{horn1}) and
(\ref{horn2}) are recovered for $a>0$ and $a=0$, respectively. From this one
see that (\ref{horn2}) is a kind of extremal case of (\ref{horn1}). In this
case, as $\rho\rightarrow+\infty$, the spacetime has a boundary of the form
$\partial M=S^{1}\times\Sigma_{3}$. Since generically, there is a smooth
singularity when $\rho\rightarrow-\infty$, it is safer to introduce two
regulators $\rho_{\pm}$, satisfying $\rho_{-}\leq\rho\leq\rho_{+}$. Due to the
fact that the base manifold has vanishing Ricci scalar, only the second term
at the r.h.s. of Eq. (\ref{variationactionabc}) remains, i.e.,%
\[
\delta I_{T}=6\kappa\beta\left[  G\ \mathcal{V}_{3}\right]  _{\rho_{-}}%
^{\rho_{+}}\ ,
\]
and it is simple to check that, since $G\left(  \rho_{-}\right)  =G\left(
\rho_{+}\right)  =0$ the boundary term (\ref{variationactionabc}) vanishes
again regardless the position of the regulators.

\bigskip

In sum, as we have shown that the black holes, wormholes and spacetime horns
are truly extrema of the action, it makes sense to evaluate the regularized
action on these solutions.

\subsection{Euclidean action for geometrically well-behaved solutions}

For the class of solutions of the form (\ref{genericabmetric}), which satisfy
(\ref{curvcond}), the bulk and boundary contributions to the regularized
action $I_{T}=I_{5}-B_{4}$, given by Eqs. (\ref{Ibulk}) and
(\ref{BoundaryTerm5}) respectively, reduce to%
\begin{align}
I_{5}  &  =\kappa\beta\left[  H\ \mathcal{I}_{3}+6\ J\ \mathcal{V}_{3}\right]
\ ,\label{genericactionevaluation}\\
B_{4}  &  =\kappa\beta\left[  h\ \mathcal{I}_{3}+6\ j\ \mathcal{V}_{3}\right]
_{\partial\Sigma}\ .\label{genericactionevaluationboundary}%
\end{align}
The functions $H$ and $J$ in the bulk term are defined by%
\begin{align}
H  &  :=-\frac{8}{l}\int AC~d\rho~,\label{ache}\\
J  &  :=\frac{4}{l^{3}}\int\left[  \left(  C^{2}\right)  ^{\prime}\left(
AC\right)  ^{\prime}-\frac{4}{3}AC^{3}\right]  \ d\rho\ ,\label{jota}%
\end{align}
where the integrals are taken along the whole range of $\rho$. For the
boundary term (\ref{genericactionevaluationboundary}), the functions $h$ and
$j$ are respectively defined by%
\begin{align}
h  &  =-\frac{2}{l}\left(  AC\right)  ^{\prime}\ ,\label{acheborde}\\
j  &  =-\frac{1}{l^{3}}\left[  \left(  AC\right)  ^{\prime}\left(  \frac
{C^{2}}{3}-C^{\prime2}\right)  +\left(  C^{2}\right)  ^{\prime}\left(
\frac{AC}{3}-A^{\prime}C^{\prime}\right)  \right]  \ .\label{jotaborde}%
\end{align}

Now it is straightforward to evaluate the regularized Euclidean action for the
class of solutions under consideration.

\bigskip

$\circ$ \emph{Black holes: }In order to obtain the regularized Euclidean
action for the black hole (\ref{Euclidean BH}) one introduces the regulator
$\rho_{a}$, such that the range of the proper radial distance is given by
$0\leq\rho\leq\rho_{a}$. The regularized action $I_{T}$ for the black hole is%
\begin{equation}
I_{T}=4\pi\kappa r_{+}\left[  \mathcal{I}_{3}+\frac{r_{+}^{2}}{l^{2}%
}\mathcal{V}_{3}\right]  \ .\label{Itbh}%
\end{equation}
Note that the action is finite and independent of the regulator $\rho_{a}$.

For a fixed temperature, the Euclidean action (\ref{Itbh}) is related to the
free energy $F$ in the canonical ensemble as%
\begin{equation}
I_{T}=-\beta F=S-\beta M\ ,
\end{equation}
so that the mass and the entropy can be obtained from%
\begin{equation}
M=-\frac{\partial I_{T}}{\partial\beta}\ ;\ S=\left(  1-\beta\frac{\partial
}{\partial\beta}\right)  I_{T}\ .
\end{equation}
In the case of a generic base manifold $\Sigma_{3}$, the thermodynamics of the
black holes in Eq. (\ref{Euclidean BH}) turns out to be qualitatively the same
as the one described in Ref. \cite{ATZ}. In the case of base manifolds of
constant curvature it agrees with previously known results.

Note that the mass of the black hole:
\begin{equation}
M=2\kappa\frac{r_{+}^{2}}{l^{2}}\left[  \mathcal{I}_{3}+\frac{3r_{+}^{2}%
}{l^{2}}\mathcal{V}_{3}\right]  \ ,\label{massactionbh}%
\end{equation}
is very sensitive to the geometry of the base manifold. For a fixed base
manifold with $\mathcal{I}_{3}<0$, $M$ is bounded from below by $M_{0}%
:=-\frac{\kappa}{6}\frac{\mathcal{I}_{3}^{2}}{\mathcal{V}_{3}}$. Note that
$M_{0}$ can be further minimized due to the freedom in the choice of the base
manifold. Even more interesting is the fact that, among the solutions with a
given base manifold satisfying $\mathcal{I}_{3}<0$, the Euclidean action
(\ref{Itbh}) has a minimum value, attained at
\begin{equation}
r_{+}=l\sqrt{\frac{-\mathcal{I}_{3}}{3\mathcal{V}_{3}}},\label{mineuc}%
\end{equation}
that can be written in terms of the Yamabe functional $Y_{3}:=\frac
{\mathcal{I}_{3}}{\mathcal{V}_{3}^{1/3}}$ \cite{yamabe}
\begin{equation}
I_{T_{0}}=-\frac{8\sqrt{3}}{9}\pi\kappa l|Y_{3}|^{3/2}\ .\label{Ito}%
\end{equation}

Note that the freedom in the choice of the boundary metric allows further
minimization of the extremum of the action (\ref{Ito}). This can be performed
by choosing $\Sigma_{3}$ as a stationary point of the Yamabe functional. Since
it is well known that the Yamabe functional has critical points for Einstein
metrics, and three-dimensional Einstein metrics are metrics of constant
curvature, the base manifold turns out to be of negative constant curvature.

\bigskip

$\circ$ \emph{Wormholes: }The Euclidean continuation of the wormhole metrics
(\ref{DOTworm}) and (\ref{worm2}) are smooth independently of the Euclidean
time period $\beta$. The Euclidean action $I_{T}=I_{5}-B_{4}$, is evaluated
introducing regulators such that $\rho_{-}\leq\rho\leq\rho_{+}$.

\bigskip

In the case of the Euclidean wormhole (\ref{DOTworm}) the regularized
Euclidean action vanishes regardless the position of the regulators, since%
\begin{equation}
I_{5}=B_{4}=2\kappa l\beta\mathcal{V}_{3}\left[  3\sinh\left(  \rho
_{0}\right)  \!+\!8\cosh^{3}\left(  \rho\right)  \sinh\left(  \rho
\!-\!\rho_{0}\right)  \right]  _{\rho_{-}}^{\rho_{+}}\ .
\end{equation}
Consequently, the mass of this spacetime also vanishes, since $M=-\frac
{\partial I_{T}}{\partial\beta}=0$.

\bigskip

For the wormhole (\ref{worm2}) the Euclidean action reads%
\begin{equation}
I_{T}=6\kappa\beta\mathcal{V}_{3}\left[  \left(  J-j\right)  -\left(
H-h\right)  \right]  \ ,
\end{equation}
with%
\begin{align}
H  &  =-2l\left.  \left(  e^{2\rho}+2\rho\right)  \right\vert _{\rho_{-}%
}^{\rho_{+}}~,\\
J  &  =-\frac{1}{3}l\left.  \left(  -e^{4\rho}+3e^{2\rho}+12\rho-e^{-2\rho
}\right)  \right\vert _{\rho_{-}}^{\rho_{+}}~,\\
h  &  =-2l\left.  e^{2\rho}\right\vert _{\rho_{-}}^{\rho_{+}}~,\nonumber\\
j  &  =-\frac{1}{3}l\left.  \left(  -e^{-4\rho}+3e^{2\rho}-e^{-2\rho}\right)
\right\vert _{\rho_{-}}^{\rho_{+}}\ .
\end{align}
The regularized action vanishes again independently of $\rho_{\pm}$, and so
does it mass.

It is worth pointing out that both wormholes can be regarded as instantons
with vanishing Euclidean action.

\bigskip

$\circ$ \emph{Spacetime horns: }The Euclidean continuation of the spacetime
horns (\ref{horn1}) and (\ref{horn2}) have arbitrary $\beta$. Let us recall
that when $\rho\rightarrow+\infty$, the spacetime has a boundary of the form
$\partial M=S^{1}\times\Sigma_{3}$, and due to the presence of the singularity
at $\rho\rightarrow-\infty$, we introduce regulators $\rho_{\pm}$, such that
$\rho_{-}\leq\rho\leq\rho_{+}$. Since the Ricci scalar of $\Sigma_{3}$
vanishes, the regularized action for the spacetime horns reduce to%
\begin{equation}
I_{T}=6\kappa\beta\mathcal{V}_{3}\left(  J-j\right)  \ .
\end{equation}

For the spacetime horn (\ref{horn1}), the Euclidean action
\begin{align}
J &  =\frac{4}{3}l\left.  \left(  e^{4\rho+\rho_{0}}-e^{2\rho-\rho_{0}%
}\right)  \right\vert _{\rho_{-}}^{\rho_{+}}\ ,\\
j &  =\frac{4}{3}l\left.  \left(  e^{4\rho+\rho_{0}}-e^{2\rho-\rho_{0}%
}\right)  \right\vert ^{\rho_{+}}\ .\nonumber
\end{align}
vanish. Note that it was necessary to take the limit $\rho_{-}\rightarrow
-\infty$.

\bigskip

In the case of the spacetime horn (\ref{horn2}), in the limit $\rho
_{-}\rightarrow-\infty$, the regularized action also vanishes since%
\begin{align}
J  &  =-\frac{8}{3}l\left.  e^{2\rho}\right\vert _{\rho_{-}}^{\rho_{+}}\ ,\\
j  &  =-\frac{8}{3}l\left.  e^{2\rho}\right\vert ^{\rho_{+}}\ .\nonumber
\end{align}

As a consequence, the masses of the spacetime horns vanishes.

\bigskip

The mass for the spacetime metrics discussed here can also be obtained from a
suitable surface integral coming from a direct application of Noether's
theorem to the regularized action functional.

\section{Mass from a surface integral}

\label{mass}

As in section IV it was shown that the geometrically well behaved solutions
are truly extrema of the regularized action, one is able to compute the mass
from the following surface integral%
\begin{equation}
Q\left(  \xi\right)  =\frac{\kappa}{l}\!\int_{\partial\Sigma}\!\epsilon
_{abcde}\left(  I_{\xi}\theta^{ab}e^{c}+\theta^{ab}I_{\xi}e^{c}\right)
\!\left(  \!\tilde{R}^{de}+\frac{1}{2}\theta_{\ f}^{d}\theta^{fe}+\frac
{1}{2l^{2}}e^{d}e^{e}\right)  ,\label{Qchi}%
\end{equation}
obtained by the straightforward application of Noether's theorem \footnote{The
action of the contraction operator $I_{\xi}$ over a $p$-form $\alpha_{p}%
=\frac{1}{p!}\alpha_{\mu_{1}\cdots\mu_{p}}dx^{\mu_{1}}\cdots dx^{\mu_{p}}$ is
given by $I_{\xi}\alpha_{p}=\frac{1}{(p-1)!}\xi^{\nu}\alpha_{\nu\mu_{1}%
\cdots\mu_{p-1}}dx^{\mu_{1}}\cdots dx^{\mu_{p-1}}$, and $\partial\Sigma$
stands for the boundary of the spacelike section.}. Here $\xi=\partial_{t}$ is
the timelike Killing vector.

For a metric of the form (\ref{genericabmetric}), satisfying (\ref{curvcond}),
(\ref{Qchi}) gives%
\begin{equation}
M=2\frac{\kappa}{l}\left[  \left(  A^{\prime}C-C^{\prime}A\right)  \left(
\mathcal{I}_{3}+\frac{3}{l^{2}}\left(  C^{2}-C^{\prime2}\right)
\mathcal{V}_{3}\right)  \right]  _{\partial\Sigma}\ ,\label{evaluatedmass}%
\end{equation}
which can be explicitly evaluated for the black holes, wormholes and spacetime horns.

\bigskip

$\circ$ \emph{Black holes: }For the black hole metric (\ref{BH-metric}) the
mass in Eq. (\ref{evaluatedmass}) reads%
\begin{equation}
M=2\kappa\frac{r_{+}^{2}}{l^{2}}\left[  \mathcal{I}_{3}+\frac{3r_{+}^{2}%
}{l^{2}}\mathcal{V}_{3}\right]  \ .
\end{equation}
It is reassuring to verify that it coincides with the mass computed within the
Euclidean approach in Eq. (\ref{massactionbh}).

\bigskip

$\circ$ \emph{Wormholes: }As explained in Ref.\cite{DOTwormhole}, for the
wormhole (\ref{DOTworm}), one obtains that the contribution to the total mass
coming from each boundary reads%
\begin{equation}
M_{\pm}=Q_{\pm}\left(  \partial_{t}\right)  =\pm6\kappa\mathcal{V}_{3}%
\sinh\left(  \rho_{0}\right)  ,
\end{equation}
where $Q_{\pm}\left(  \partial_{t}\right)  $ is the value of (\ref{Qchi}) at
$\partial\Sigma_{\pm}$, which again does not depend on $\rho_{+}$ and
$\rho_{-}$. The opposite signs of $M_{\pm}$, are due to the fact that the
boundaries of the spatial section have opposite orientation. The integration
constant $\rho_{0}$ can be regarded as a parameter for the apparent mass at
each side of the wormhole, which vanishes only when the solution acquires
reflection symmetry, i.e., for $\rho_{0}=0$. This means that for a positive
value of $\rho_{0}$, the mass of the wormhole appears to be positive for
observers located at $\rho_{+}$, and negative for the ones at $\rho_{-}$, with
a vanishing total mass $M=M_{+}+M_{-}=0$.

\bigskip

For the wormhole (\ref{worm2}) the total mass also vanishes since the
contribution to the surface integral (\ref{Qchi}) coming from each boundary
reads%
\begin{equation}
M_{\pm}=\mp\ 6\kappa\mathcal{V}_{3}\ ,
\end{equation}
so that $M=M_{+}+M_{-}=0$.

Note that $M_{\pm}$ are concrete examples of Wheeler's conception of
\emph{\textquotedblleft mass without mass"}.

\bigskip

$\circ$ \emph{Spacetime horns: }For the spacetime horns (\ref{horn1}) and
(\ref{horn2}) the masses also vanish. This can be easily verified from
(\ref{evaluatedmass}), the fact that $\mathcal{I}_{3}=0$ (since $\tilde{R}%
=0$), and that the warp factor of the base manifold, $C=e^{\rho}$, satisfies
$\left(  C^{2}-C^{\prime2}\right)  =0$.

\section{Discussion and comments}

\label{conc}

An exhaustive classification for the class of metrics (\ref{Ansatz}) which are
solutions of the Einstein-Gauss-Bonnet theory in five dimensions has been
performed. In Section II, it was shown that for generic values of the coupling
constants, the base manifold $\Sigma_{3}$ must be necessarily of constant
curvature, and consequently, the solution reduces to the topological extension
of the Boulware-Deser metric, for which $f^{2}=g^{2}$ is given by
(\ref{gcuadradogenerico}). It has also been shown that the base manifold
admits a wider class of geometries for those special theories for which the
Gauss-Bonnet coupling acquires a precise relation in terms of the cosmological
and Newton constants, given by (\ref{tuneo}).

Remarkably, the additional freedom in the choice of the metric at the
boundary, which determines $\Sigma_{3}$, allows the existence of three main
branches of geometries in the bulk (Section II). \newline The geometrically
well-behaved metrics among this class correspond to the case of negative
cosmological constant.

If the boundary metric is chosen to be such that $\Sigma_{3}$ is an arbitrary,
but fixed, base manifold, the solution is given by (\ref{BH-metric}), and
describes black holes whose horizon geometry inherits the metric of the base
manifold. These solutions generalize those in \cite{Cai-Soh} and \cite{ATZ},
for which $\Sigma_{3}$ was assumed to be of constant curvature, which, in the
case of spherical symmetry, reduce to the metrics in \cite{BD}, \cite{BTZ}.

If the metric at the boundary is chosen so that the base manifold $\Sigma_{3}
$ possesses a constant negative Ricci scalar, two different kinds of wormhole
solutions in vacuum are obtained. One of them, given in (\ref{DOTworm}), was
found previously in \cite{DOTwormhole} and describes a wormhole connecting two
asymptotic regions whose metrics approach that of AdS spacetime, but with a
different base manifold. The other solution, given in (\ref{worm2}), describes
a brand new wormhole connecting an asymptotically locally AdS spacetime at one
side of the throat, with a nontrivial curved and smooth spacetime on the other
side. Note that, in view of Yamabe's theorem \cite{yamabe}, any compact
Riemannian manifold has a conformally related Riemannian metric with constant
Ricci scalar, so that there are many possible choices for $\Sigma_{3}$.

For boundary metrics for which the base manifold $\Sigma_{3}$ has vanishing
Ricci scalar, a different class of solutions is shown to exist. For these
\textquotedblleft spacetime horns" the warp factor of the base manifold is an
exponential of the proper radial distance, and generically possess a
singularity as $\rho\rightarrow-\infty$. As explained in Sec. III, this
singularity is weaker than that of the five-dimensional Schwarzschild solution
with negative mass, and it is also weaker than that of a conifold.

It has also been shown that if $\Sigma_{3}$ is of constant curvature, due to
certain class of degeneration of the field equations for the theories
satisfying (\ref{tuneo}), there is a special case where the metric admits an
arbitrary redshift function. This degeneracy is a known feature of the class
of theories considered here \cite{dege}. A similar degeneracy has been found
in the context of Birkhoff's theorem for the Einstein-Gauss-Bonnet theory
\cite{Charmousis-Dufaux}, \cite{Zegers}, which cannot be removed by a
coordinate transformation \cite{Deser}. Birkhoff's theorem has also been
discussed in the context of theories contaning a dilaton and an axion field
coupled with a Gauss-Bonnet term in \cite{ACD}.

In the sense of the AdS/CFT correspondence \cite{magoo}, the dual CFT living
at the boundary, which in our case is of the form $S^{1}\times\Sigma_{3}$,
should acquire a radically different behavior according to the choice of
$\Sigma_{3}$, since it has been shown that the bulk metric turns out to be
very sensitive to the geometry of the base manifold. Notice that the existence
of asymptotically AdS wormholes raises some puzzles concerning the AdS/CFT
conjecture \cite{WY}, \cite{MM}, \cite{AOP}.

It is worth pointing out that an interesting effect occurs for geodesics with
angular momentum for the generic class of spacetimes given by
(\ref{genericmetricrecu}), among which the wormholes and spacetime horns are
included. In a few words, there are regions for which the effective potential
cannot have a minimum, since the gravitational force points in the same
direction as the centrifugal force. Therefore, within these regions, there is
at most a single turning point, and consequently bounded orbits cannot exist.

In Sec. IV, it was shown that the geometrically well-behaved solutions have
finite Euclidean action. In the case of black holes, the Euclidean action
reduces to the free energy in the canonical ensemble. It has also been shown
that black holes whose base manifolds are such that its Einstein-Hilbert
action $\mathcal{I}_{3}$ is negative, have a nontrivial ground state, for
which its Euclidean action is an increasing function of the Yamabe functional,
and therefore, its value is further extremized when the base manifold
$\Sigma_{3}$ is of constant curvature.

In the case of wormholes, the Euclidean continuation is regular for an
arbitrary Euclidean time period $\beta$, and they can be regarded as
instantons with vanishing Euclidean action and mass. For the spacetime horns,
their regularized action and mass vanish; so that in this sense, the
singularity is as tractable as it is for a vortex.

\bigskip

It is simple to see that, the class of solutions discussed here can be
embedded into the locally supersymmetric extension of the five-dimensional
Einstein-Gauss-Bonnet for the choice of coefficients (\ref{tuneo})
\cite{CHAM2}, \cite{sugraricalld}. As a consequence, the black holes
(\ref{BH-metric}) admit a ground state with unbroken supersymmetries whose
Killing spinors were explicitly obtained in \cite{KS}. In this case the base
manifold must necessarily be Einstein.

For the special coefficients (\ref{tuneo}), the freedom in the choice of the
base manifold allows to consider as a particular case, base manifolds of the
form $\Sigma_{3}=S^{1}\times\Sigma_{2}$. This can be performed for all the
branches, but not for the degenerate one. This means that compactification to
four dimensions for the black holes (\ref{BH-metric}), the wormholes
(\ref{DOTworm}), (\ref{worm2}), and the spacetime horns (\ref{horn1}),
(\ref{horn2}) is straightforward. Therefore, the dimensionally reduced
solutions posses the same causal behavior as their five-dimensional seeds, but
they are supported by a nontrivial dilaton field with a nonvanishing
stress-energy tensor. Further compactifications have been found in Refs.
\cite{SUPERMTZ}, \cite{GOT} and \cite{DBS}. The dimensional reduction of the
Einstein-Gauss-Bonnet theory has been discussed in Ref. \cite{Muller}, and for
the special choice of coefficients (\ref{tuneo}), it has been discussed
recently in Ref. \cite{ARZ}, including new exact solutions.

\bigskip

For the Einstein-Gauss-Bonnet theory, black holes with nontrivial horizon
geometry have also been discussed in Refs. \cite{CO}, \cite{Hideki}; it is
worth pointing out that the stability of Gauss-Bonnet black holes is fairly
different than that of the Schwarszchild solution \cite{DG1}, \cite{Neupane},
\cite{DG2}, \cite{DG3}, \cite{BDG}. Solutions possessing NUT charge have been
found in \cite{DeMann}. Wormhole solutions for this theory, in the presence of
matter that does not violate the weak energy condition have been shown to
exist provided the Gauss-Bonnet coupling constant is negative and bounded
according to the shape of the solution \cite{BKar}. Thin shells wormholes for
this theory have been discussed recently in \cite{TSE}. For the pure
Gauss-Bonnet theory, i.e., for the action (\ref{action}) with $\alpha
_{1}=\alpha_{0}=0$, wormhole solutions in vacuum, for which there is a jump in
the extrinsic curvature along a spacelike surface, have been shown to exist
recently \cite{WG}. Higher dimensional wormhole solutions have also been
discussed in the context of braneworlds, see e.g., \cite{Lobo} and references therein.

\bigskip

As a final remark, it is worth pointing out that the results found here are
not peculiarities of five-dimensional gravity, and similar structures can be
found in higher dimensional spacetimes \cite{DOTBasehigh}.

\bigskip

\textit{Acknowledgments.-- }We thank Arturo G\'{o}mez for thorough reading of
this paper and for useful remarks. G.D is supported by CONICET. J.O. thank the
support of projects MECESUP UCO-0209 and MECESUP USA-0108. J. O. and R. T.
thanks the organizers of \textquotedblleft Grav06, Fifty years of FaMAF \&
Workshop on Global Problems in GR", held in C\'{o}rdoba, for their warm
hospitality. This work was partially funded by FONDECYT grants 1040921,
1051056, 1061291, 1071125; Secyt-UNC and CONICET. This work was funded by an
institutional grant to CECS of the Millennium Science Initiative, Chile and
also benefits from the generous support to CECS by Empresas CMPC.

\end{document}